\documentstyle[amssymb,aps,pra,twocolumn]{revtex}
\begin{document}
\title{A possible cooling effect in high temperature superconductors }
\author{Anatoly A.~Svidzinsky}
\address{Department of Physics, Stanford University, Stanford, CA 94305-4060}
\date{\today }
\maketitle

\begin{abstract}
We show that an adiabatic increase of the supercurrent along a
superconductor with lines of nodes of the order parameter on the Fermi
surface can result in a cooling effect. The maximum cooling occurs if the
supercurrent increases up to its critical value. The effect can also be
observed in a mixed state of a bulk sample. An estimate of the energy
dissipation shows that substantial cooling can be performed during a
reasonable time even in the microkelvin regime.
\end{abstract}


Although the mechanism of high temperature superconductivity still remains
unclear there is strong evidence that the order parameter in high
temperature superconductors has nodes on the Fermi surface \cite{Tsue00}.
Such a property is also peculiar to some superconductors with heavy
fermions. In this paper we show that presence of nodes of the order
parameter can lead to a cooling effect. The cooling is reached by
adiabatically increasing the supercurrent around a superconducting ring or a
cylinder.

Currently the lowest temperature of a solid ($T\sim 1\mu K$) is achieved by
using a method of adiabatic nuclear demagnetization \cite{WHRP97,WHRP}.
Spontaneous magnetic ordering of the nuclear magnetic moments represents the
low temperature limit for nuclear refrigeration. The ordering temperature
due to nuclear dipole-dipole interaction is typically a fraction of a
microkelvin and, therefore, to achieve temperatures lower than $0.1\mu K$ a
new refrigeration technology is needed. In our cooling mechanism the
conduction electrons are cooled during the adiabatic increase of the
supercurrent instead of the nuclear spin system being cooled in the nuclear
demagnetization process. Further development of the idea of cooling using
the phenomena of superconductivity (superfluidity) could be promising to
achieve the lowest temperature of solids.

\begin{center}
{\bf Cooling effect in clean superconductors}
\end{center}

Let us consider a superconducting ring (or a cylinder) made from a clean
high temperature superconductor. For estimates one can assume the Fermi
surface to be a cylinder and the order parameter has a simple $d$-wave form $%
\Delta ({\bbox p})=\Delta _0\cos (2\phi )$, where $\phi $ is the polar angle
in the $ab$ crystalline plane. Taking into account the real form of the
Fermi surface might change numerical coefficients, but these details are not
necessary for estimating the magnitude of the cooling effect. Let the $c$
-axis of the superconductor be parallel to the symmetry axis of the ring and
the temperature of the system is $T\ll T_c$, where $T_c$ is the
superconducting transition temperature.

At low temperatures the main contribution to the electronic specific heat of
the superconductor arises from quasiparticles that are activated above the
gap in narrow vicinities of the nodes of the order parameter. The energy of
these quasiparticles is given by 
\begin{equation}
\label{1}E({\bbox p})=\sqrt{\xi ({\bbox p})^2+|\Delta ({\bbox p})|^2}, 
\end{equation}
and the electronic specific heat is \cite{BS}: 
\begin{equation}
\label{2}C(T)=36\zeta (3)N_F\frac{T^2}{\Delta _0}, 
\end{equation}
where $N_F$ is the density of states at the Fermi surface for normal metal
and $\zeta (3)\simeq 1.202$ is the Riemann zeta function (throughout this
paper, apart from formula (\ref{22}), we use units in which $k_B=1$).
Therefore, the entropy of the system has the form: 
\begin{equation}
\label{3}S(T)=\int_0^T\frac{C(T)}TdT=18\zeta (3)N_F\frac{T^2}{\Delta _0}. 
\end{equation}

Let us now consider the same superconducting ring with a supercurrent
uniformly distributed across the cross section of the ring. This situation
is possible if the thickness of the ring $d$ is less than penetration depth
of the magnetic field $\lambda $: $d\lesssim \lambda $. In the presence of
such superflow with a superfluid velocity ${\bbox v}_s$, the quasiparticles
can be considered quasiclassically with the energy \cite{XYS,BS}: 
\begin{equation}
\label{4}E({\bbox p,\bbox v}_s)=\sqrt{\xi ({\bbox p})^2+|\Delta ({\bbox p}%
)|^2}+{\bbox p}_{F\perp }\cdot {\bbox v}_s, 
\end{equation}
where ${\bbox p}_{F\perp }$ is the component of the Fermi momentum in the
direction perpendicular to the $c$-axis. Due to the existence of the
supercurrent, quasiparticle states with negative energy appear for momentum
directions within the vicinity of nodes of the order parameter. These states
are occupied even at zero temperature. The presence of these states gives a
term, linear in $T$, in the low temperature specific heat, which is dominant
if $p_{F\perp }v_s\gg T$ \cite{BS}. After a space averaging over the ring
this term can be written in the form $C(T)=4\pi N_FTp_{F\perp }v_s/3\Delta
_0 $. As a result, the entropy in the presence of the supercurrent is: 
\begin{equation}
\label{6}S(T)=\frac 43\pi N_F\frac{p_{F\perp }v_s}{\Delta _0}T. 
\end{equation}

Let us assume that at an initial moment of time there is no superflow along
the ring, and the ring is thermally isolated. A magnetic field is then
turned on adiabatically so that the magnetic flux through the ring $\Phi $
slowly increases in time. The induced electric field $E\sim \dot \Phi /lc$ ($%
l$ is the ring's length) generates current inside the ring which consists,
in general, of superconducting and dissipative parts. The dissipative part
of the current $I_{diss}$ is proportional to $E$, and the energy dissipation
can be estimated as 
\begin{equation}
\label{8}\int_{t=0}^{t_f}I_{diss}Eldt\propto \int_{t=0}^{t_f}\dot \Phi
^2dt\sim \frac{\Phi _f^2}{t_f}. 
\end{equation}
The dissipation can be made negligibly small if the time $t_f$ to build up
the magnetic field is big enough. In this process all the work which is
produced by the induced electric field goes to increase the supercurrent
around the ring; and the part of this work that dissipates into heat and
changes the entropy can be made negligibly small. The necessary cooling time 
$t_f$ is estimated at the last section of this article.

The equation of entropy conservation in the adiabatic process of increasing
the supercurrent has the form:

\begin{equation}
\label{9}18\zeta (3)N_F\frac{T_0^2}{\Delta _0}=\frac 43\pi N_F\frac{%
p_{F\perp }v_s}{\Delta _0}T_f, 
\end{equation}
where $T_0$ and $T_f$ are the initial and the final temperatures of the
ring. As a result, the final temperature $T_f$ after the adiabatic increase
of the supercurrent is 
\begin{equation}
\label{11}T_f=\frac{27\zeta (3)}{2\pi }\frac{T_0^2}{p_{F\perp }v_s}=5.2\frac{%
T_0^2}{p_{F\perp }v_s} 
\end{equation}
This formula is applicable if $p_{F\perp }v_s\gg T_f$. The maximum cooling
occurs when the supercurrent (and the superfluid velocity) is equal to its
critical value, that is $p_{F\perp }v_s\sim 2\Delta _0=4.28T_c$ and 
\begin{equation}
\label{12}T_{f\min }\approx \frac{T_0^2}{T_c} 
\end{equation}
So, to achieve substantial cooling, it is important to use a superconductor
with high $T_c$. If, for example, $T_0=1K$ and $T_c=100K$, after the
adiabatic increase of the supercurrent, the ring cools down to $T_f=0.01K$.

One should mention that we base our estimates on the assumption of uniform
temperature distribution during the cooling process. However, the weight of
quasiparticle states shifted from above the Fermi level to below depends on
the angle the supercurrent makes with the nodal directions, which changes
around the ring. This would provide a larger cooling effect at four points
on the ring (at which ${\bbox v}_s$ constitutes $45^{\circ }$ angle with
respect to $a$, $b$ axes) compared to elsewhere and in the ideal case of
zero thermoconductivity the temperature distribution over the ring would be
nonuniform with $T_{max}/T_{min}=\sqrt{2}$. However, finite
thermoconductivity equalizes the temperature around the ring and nonuniform
cooling would be very difficult to observe experimentally. Estimates show
that nonuniform cooling can be detected if the cooling time is less than
about $R^2/v_F^2\tau _N$, where $R$ is the radius of the ring, $v_F$ is the
Fermi velocity and $\tau _N$ is the relaxation time in the normal state. For
reasonable parameters we obtain that the necessary cooling time should be of
the order of nanosecond. Real cooling time is much larger than nanosecond
which guarantees approximately uniform temperature distribution during the
cooling process. As a result the entropy dissipation due to the thermal
current is negligible and does not modify our formulas.

Also we do not consider here the effect of fluxoid quantization which is
negligible when the flux through the ring is much greater than the flux
quantum. This gives a restriction on the superfluid velocity $v_s\gg v_Fa/R$%
, where $a$ is the interatomic spacing. From the other hand, the critical
superfluid velocity is of the order of $2\Delta _0/p_{F\perp }\approx
v_Fa/\xi $, where $\xi $ is the coherence length. So, we can omit the effect
of fluxoid quantization in our problem if $R\gg \xi $.

One should also note that value $v_s\sim 2\Delta _0/p_{F\perp }$ determines
depairing current density $j_{depair}$. Because of the grain structure of
high temperature superconductors the critical current density $j_c$ is
usually much less than the depairing value. A misalignment of adjacent
grains suppresses the order parameter at the grain boundary and results in
Josephson tunneling behavior and a near-exponential decrease in $j_c$ with
misorientation angle. If the current through the superconductor is increased
to $j_c$, rather than $j_{depair}$, the estimate (\ref{12}) for the cooling
temperature should be multiplied by the factor $j_{depair}/j_c$. To reach $%
j_c$ close to the depairing value one should use materials with well-aligned
grains, which can be achieved in thin films \cite{Shea94}. For example, $%
YBa_2Cu_3O_{7-\delta }$ films were obtained with the critical current
density $j_c\approx 40MA/cm^2$, so that $j_{depair}/j_c\approx 5$ \cite
{Vere01}. It is worth to note that if the cooling is achieved by applying
external magnetic field, rather than creating a supercurrent along the film,
the grain structure would not be an obstacle. In such a way, the magnetic
field induces screening supercurrent in each grain. The intragrain
supercurrent is not limited by the small value of the Josephson critical
current in weak links between the grains and can be close to $j_{depair}$.

It is interesting to estimate the value of the cooling effect for a bulk
sample in a mixed state. In a magnetic field $H$ ($H_{c1},H_{c2}T^2/T_c^2\ll
H\ll H_{c2}$) the entropy of a clean superconductor with line(s) of nodes is 
$S\propto T\sqrt{H/H_{c2}}$ \cite{V93}. So, an adiabatic increase of
magnetic field from $H_0$ to $H_f$ cools down the sample to the temperature 
\begin{equation}
\label{a12}T_f=T_0\sqrt{\frac{H_0}{H_f}}. 
\end{equation}

One should note that in conventional $s$ -wave superconductors the
electronic contribution to the entropy is exponentially small (due to finite
gap in the excitation spectrum) and thus not dominant at low temperatures.
However, in a mixed state of $s$ -wave superconductors normal regions in the
vortex cores give the linear temperature contribution to the electron
entropy $S\sim N_FTH/H_{c2}$, which can exceed the phonon contribution at
low enough temperature $T$ (typically less then $1$K) 
\begin{equation}
\label{b12}T\lesssim \frac{T_D}4\left( \frac{T_D}{T_F}\right) ^{1/2}\left( 
\frac H{H_{c2}}\right) ^{1/2}, 
\end{equation}
where $T_D$ and $T_F$ are Debye and Fermi temperatures accordingly. In this
temperature region an adiabatic increase of applied magnetic field $H$ ($%
H_{c1}<H<H_{c2}$) results in the cooling effect $T_f=T_0H_0/H_f$.

\begin{center}
{\bf Effect of impurities}
\end{center}

At very low temperatures (when $T\lesssim\gamma$) the effect of
impurity-induced bound states becomes significant. Let us consider the
cooling effect at temperatures when the regime $T\lesssim\gamma \ll T_c$ is
achieved. The energy scale $\gamma $ is the bandwidth of quasiparticle
states bound to impurities \cite{S,HVW}. For an order parameter with a line
of nodes, the bandwidth $\gamma $ and density of the quasiparticle states at
zero energy $N(0)$ are finite for any nonzero concentration of impurities.
The quantities $\gamma $ and $N(0)$ depend on the impurity concentration $%
n_i $ and the scattering phase shift $\delta _0$. For example, for
scattering in the unitary limit ($\delta _0=\pi /2$) and the $d$-wave order
parameter, $\gamma \sim \sqrt{\pi \Gamma _u\Delta _0/2 \text{ }}$, $%
N(0)=8\gamma N_F\log \left( 4\Delta _0/\gamma \right) /\pi \Delta _0$, where 
$\Gamma _u=n_i/\pi N_F$. In the opposite limit (Born approximation with $%
\delta _0\rightarrow 0$) the bandwidth $\gamma $ is exponentially small $%
\gamma \sim 4\Delta _0\exp (-\pi \Delta _0/2\Gamma _u\delta _0^2)$ and,
therefore, the regime $T\lesssim\gamma $ can be very difficult to achieve
experimentally. For most high temperature and heavy fermion superconductors
the impurity scattering is strong (or an intermediate strength) rather than
in the Born limit \cite{T97,S98,PP86,SMV}.

For $T\lesssim \gamma $ and in the absence of superflow the electronic
specific heat is given by $C(T)=\pi ^2N(0)T/3$. So, instead of (\ref{3}),
the entropy of the system is 
\begin{equation}
\label{13}S(T)=\frac{\pi ^2}3N(0)T. 
\end{equation}
After the adiabatic process of creating the supercurrent around the ring
with $p_{F\perp }v_s\gg \gamma $, the entropy of the system is given
approximately by the same expression (\ref{6}) as for the clean
superconductor. As a result, the equation of the entropy conservation gives
the following formula for the final temperature: 
\begin{equation}
\label{15}T_f=\frac \pi 4\frac{N(0)}{N_F}\frac{\Delta _0}{p_{F\perp }v_s}T_0 
\end{equation}
For maximum cooling ($p_{F\perp }v_s\sim 2\Delta _0$) 
\begin{equation}
\label{16}T_{f\min }\approx \frac \pi 8\frac{N(0)}{N_F}T_0 
\end{equation}
For scattering in the unitary limit we get: 
\begin{equation}
\label{17}T_{f\min }\approx \frac \gamma {\Delta _0}\log \left( \frac{%
4\Delta _0}\gamma \right) T_0\sim \frac \gamma {T_c}T_0 
\end{equation}

In the case of a bulk superconductor in a mixed state the presence of
impurities results in appearance of a crossover field $H^{*}$ when
square-root magnetic field dependence of the specific heat changes to a
dependence of the form $H\ln H$ \cite{BMS}. If $H>H^{*}$ the influence of
impurities is negligible and the cooling effect is given by Eq. (\ref{a12}).
However, if $H_0<H^{*}<H_f$ then $T_f\approx T_0\sqrt{H^{*}/H_f}$.

\begin{center}
{\bf Estimation of the energy dissipation }
\end{center}

If there is a time-dependent current in a superconductor, the normal
electrons give a finite amount of dissipation because the supercurrent is
not a zero-impedance shunt at nonzero frequency $\omega $ (in a two-fluid
model $Im\sigma =n_se^2/m\omega $). This dissipation increases the entropy
of the system. Given an imposed ac current density $j=j_0\sin \left( \omega
t\right) $, the power dissipated per unit volume at low frequencies is \cite
{T}: 
\begin{equation}
\label{18}\frac{dW_{diss}}{dt}=Re\left( \frac 1\sigma \right) j^2\approx 
\frac{Re\sigma }{\left( Im\sigma \right) ^2}j^2=\frac{16\pi ^2\lambda ^4}{%
c^4 }\omega ^2Re\sigma \,j^2, 
\end{equation}
where $\sigma $ is the conductivity of the superconductor, $\lambda $ is the
penetration depth. To estimate the change of the entropy $\Delta S$ in the
process of increasing the supercurrent from $j=0$ to $j=j_0$ one can
integrate Eq. (\ref{18}) in the limits $0<t<t_f=\pi /2\omega $. Generally
speaking, the conductivity $\sigma $ can depend on the magnitude of the
supercurrent in the superconductor. However for estimation one can take $%
\sigma $ to be equal to its maximum value (at a maximum value of the
current). As a result, we obtain the following estimate for the change of
the entropy per unit volume: 
$$
\Delta S=\int_0^{t_f}\frac 1T\frac{dW_{diss}}{dt}dt\approx 
$$
\begin{equation}
\label{19}\approx \frac{16\pi ^2\lambda ^4\omega ^2}{c^4T_f}\int_0^{t_f}\sin
\left( \omega t\right) Re\sigma \,j^2dt\approx \frac{16\pi ^3\lambda ^4}{
3T_ft_fc^4}Re\sigma \,j_0^2. 
\end{equation}
The energy dissipation gives rise to a heating of the superconductor and
changes the final temperature $T_f$. To estimate the magnitude of this
effect one should add $\Delta S$ to the left side of Eq. (\ref{9}). It is
possible to obtain substantial cooling even if $\Delta S$ is larger than the
initial entropy. However, to make the cooling effective, one should increase
the supercurrent slowly enough to ensure that the dissipation does not
significantly change the entropy: $\Delta S\lesssim 4\pi N_FT_fp_{F\perp
}v_s/3\Delta _0$. Taking into account $j_0\approx en_sv_s=mc^2v_s/4\pi
e\lambda ^2$, we see that dissipation can be omitted if the time of
increasing the supercurrent from zero to its critical value satisfies the
following condition 
\begin{equation}
\label{20}t_f\geq \frac{Re\sigma }{2e^2N_Fv_F^2}\frac{\Delta _0^2}{T_f^2}. 
\end{equation}
The electrical conductivity for a superconductor with an order parameter
that vanishes along a line of nodes has a universal limiting value which is
independent of the concentration of impurities and the scattering phase
shift \cite{L}. For the two dimensional $d$-wave order parameter and an
isotropic 2D Fermi surface the universal value has the form \cite{L,HPS}: $%
Re\sigma \left( \omega \rightarrow 0,T\ll \gamma \right) \simeq
e^2N_Fv_F^2\tau _\Delta $, where $\tau _\Delta \simeq \hbar /\pi \Delta _0$
is a universal transport time. However, the superflow modifies this result
and necessarily increases the conductivity in the region interesting for
cooling (i.e. when $p_{F\perp }v_s\gg \gamma $).

To estimate the electrical conductivity in the presence of the superflow,
one can use the same semiclassical approach as for the case of thermal
conductivity \cite{BS98}. At $\omega \rightarrow 0$ and $\gamma \ll
p_{F\perp }v_s$ the final result is given by (cf. \cite{H89}, \cite{HPS}): 
$$
Re\sigma _{ij}=\frac{ne^2}{2mT}\int_{-\infty }^{+\infty }d\tilde \omega 
\frac{\tau (\tilde \omega ,v_s)}{\cosh ^2\left( \frac{\tilde \omega }{2T}%
\right) } 
$$
\begin{equation}
\label{21}\int \frac{d\phi }{2\pi }\hat p_i\hat p_j\frac{|\tilde \omega -{\ 
\bbox p}_{F\perp }{\bbox v}_s|}{\sqrt{\left( \tilde \omega -{\bbox p}%
_{F\perp }{\bbox v}_s\right) ^2-|\Delta (\phi )|^2}}, 
\end{equation}
where $\tau (\tilde \omega ,v_s)$ is the relaxation time for quasiparticle
scattering on impurities in the superconducting state, $n$ is the electron
density. For the $d$-wave order parameter we obtain for $T\ll p_{F\perp
}v_s\ll \Delta _0$, $Re\sigma _{ii}=ne^2\tau _N/m$, in the Born
approximation; $Re\sigma _{ii}=ne^2\tau _N\left( 1+|\cos \phi _i|\right)
p_{F\perp }^2v_s^2/4m\Delta _0^2$, in the unitary limit, where $\tau _N$ is
the relaxation time in the normal state, $\phi _i$ is the angle between ${%
\bbox v}_s$ and the $i$-axis, $i=a$, $b$. For the supercurrent close to its
critical value we get the estimate $Re\sigma \sim ne^2\tau
_N/m=e^2N_Fv_F^2\tau _N$ which is $\tau _N/\tau _\Delta \sim \Delta
_0^2/\gamma ^2\gg 1$ times larger than the universal value. If we substitute
this estimate into (\ref{20}), we obtain the following limitation for the
necessary cooling time 
\begin{equation}
\label{a22}t_f\geq \tau _NT_c^2/T_f^2. 
\end{equation}
In the unitary limit $\tau _N=\pi \hbar \Delta _0/4k_B\gamma ^2\sim \hbar
T_c/k_B\gamma ^2$ and the limitation can be rewritten as 
\begin{equation}
\label{22}t_f\geq \frac{T_c^2}{\gamma ^2}\frac{\hbar T_c}{k_BT_f^2}. 
\end{equation}
If, for example, $T_c\sim 100K$, $\gamma =0.01T_c$, $T_0=0.01K$, $T_f\sim
10^{-4}K$ the estimate for the necessary cooling time is $t_f\geq 12\min $.
In principle, the cooling time can be reasonable even in a submicrokelvin
region. In this region unconventional superconductors with low $T_{c\text{ }%
} $ like heavy fermion systems can be used. One of the possible candidates
is $UPt_3$, which has an order parameter with a line of nodes and $%
T_c\approx 0.5K$. If, for example, $T_c=0.5K$, $\gamma =0.1T_c$, $%
T_0=10^{-6}K$, $T_f\sim 10^{-7}K$ then from Eq. (\ref{22}) we obtain $%
t_f\geq 10h$. One should note, however, that applicability of the method is
limited by ultra low temperatures. At microkelvin temperatures the entropy
of the quasiparticles near nodes is small as compared to the nuclear
entropies and an estimate of the Korringa constant for $UPt_3$ from the
value of pure $Pt$ ($K=30mKs$) shows that in $10$ hours the electrons are
not thermally disconnected from the nuclear system.

To make effective refrigerating storage based on the method discussed here,
it is not sufficient to use a single ring or a cylinder because the allowed
width of the ring is confined by the penetration depth $\lambda $, which is
usually a fraction of a micron. Instead of a single ring, one can make a
coil from a superconducting wire or a tape so that the width of the tape is
of the order of $\lambda $, but the total width of the coil can be made much
larger. The superconducting current in the tape can be created by applying
an external magnetic field to the coil or by applying voltage to the tape
ends. Another possibility to obtain a cooling effect is to use a bulk sample
in a mixed state.

In conclusion, we have shown that an adiabatic increase of the supercurrent
through a superconductor with an order parameter with line(s) of nodes on
the Fermi surface results in a cooling effect. For maximum cooling, the
initial temperature $T_0$ decreases to $T_0^2/T_c$ for clean superconductors
and to $T_0\gamma /T_c$ for $T_0<\gamma $. The effect can also occur in a
mixed state of a bulk sample. To observe the cooling effect it is important
to be in the low temperature region in which conduction electrons give the
main contribution to the system's entropy.

I would like to thank A. Fetter and K. Moler for useful discussions. This
work was partially supported by the National Science Foundation, Grant No.
DMR 99-71518, and by Stanford University.

\end{document}